# Variational approach to unsupervised learning

Swapnil Nitin Shah (swapnil.nitin.shah@gmail.com)
IBM Systems, 11400 Burnet Road, Austin, Texas, 78727

**Deep belief networks are used extensively for unsupervised stochastic learning on large datasets. Compared to other deep learning approaches their layer-by-layer learning makes them highly scalable[5,6]. Unfortunately, the principles by which they achieve efficient learning are not well understood. Numerous attempts have been made to explain their efficiency and applicability to a wide class of learning problems in terms of principles drawn from cognitive psychology, statistics, information theory, and more recently physics[7], but quite often these imported principles lack strong scientific foundation. Here we demonstrate how one can arrive at convolutional deep belief networks as potential solution to unsupervised learning problems without making assumptions about the underlying framework. To do this, we exploit the notion of symmetry that is fundamental in machine learning, physics and other fields, utilizing the particular form of the functional renormalization group in physics.**

Unsupervised learning is a form of machine learning that learns from unclassified and uncategorized data by identifying similarities between samples of data which are provided during a training phase. It is then required to react based on whether or not a test data sample, not presented to it in the training phase, bears any of the identified similarities with the training samples[1]. Over the years, numerous approaches have been proposed and used for unsupervised learning problems to a varying degree of success. These primarily include latent variable models, network models and clustering algorithms[2]. While each of these models works well with a subset of the problems, general applicability of the solution they provide, and its evaluation thereof lacks strong grounding in scientific literature[3,4].

One model in particular, the Deep Belief Network (DBN), has received widespread attention from the Machine Learning community in recent times[5,6] owing to fast training while ensuring good fidelity of outcomes across a large subset of problems. DBNs are neural network models comprising of multiple layers of latent (hidden) variables with each pair of consecutive layers along with the connections between them constituting a Restricted Boltzmann Machine (RBM). When trained on samples of data, a DBN learns the probability distribution of the samples.

Mehta and Schwab[7] provide an analogy between DBNs and the mathematical tool of variational Renormalization Group (RG) which is widely used to study physical systems at various energy scales. It shows how a step of real space renormalization of spin systems is analogous to training of an RBM. The hidden spins of the RBM serve as coarse grained description of the visible spins. The renormalization step, when done exactly, ensures that the learnt variational joint distribution of hidden and visible spins when marginalized over the hidden spins is exactly equal to the prior probability of the visible spins. Albeit an important first step towards providing evidence in favor of DBNs, a more fundamental understanding of the underlying concepts is required to explain (a) how DBNs provide a solution to the general stochastic unsupervised learning problem, (b) why there is an explicit need for stacking RBMs (in a DBN) as against use of single RBM, (c) criteria to ensure that trained RBM produces exact RG transformation and (d) applicability to general problems/systems that cannot be reduced to the Ising model or which have non-thermal fixed points (continuous fourier spectra). In this paper, we attempt to address these questions after having introduced the requisite mathematical formalism.

Given a set $X$ and a structure $S(x)$ $x \epsilon X$, a transformation $f: X \rightarrow X$ is a symmetry if $f$ is invertible and preserves $S$. The group of all such transformations is called the Symmetry group of $S$ with function composition as the group operation. Symmetry can be a discrete transformation or a continuous one. The transformation group of latter is what is known as a Lie group[8]. In this context, unsupervised learning can be thought of as learning the structure $S$ which is invariant under the symmetries of the data.

Let a data-point be described by a field $\phi \epsilon \Phi$ over the $d$ data dimensions and a functional $S[\phi]: \Phi \rightarrow \mathbb{R}$ where $\mathbb{R}$ is the set of reals. A continuous infinitesimal transformation then has the following form

$$\phi'(x) = \phi(x) + \epsilon F(x,\phi), \quad \epsilon \to 0 \qquad (1)$$

The functional S varies as

$$S[\phi'] = S[\phi] + \epsilon \int d^d x \frac{\delta S}{\delta \phi} \cdot F(x,\phi) \qquad (2)$$

If the transformation (1) is a symmetry of $S$, we have

$$S[\phi'] = S[\phi] \;\Rightarrow\; \int d^d x \frac{\delta S}{\delta \phi} \cdot F(x,\phi) = 0 \qquad (3)$$

If one considers a probability distribution $P[\phi]$ over all field configurations from which the data is sampled, the stochastic unsupervised learning problem now becomes learning a structure $\Pi$ which remains invariant under symmetries of the average field $\phi_a(x)$, defined as the functional integral

$$\phi_a(x) = \langle \phi(x) \rangle = \int [D\phi]\, P[\phi] \cdot \phi(x) \qquad (4)$$

where $[D\phi]$ is the functional integral measure. Hence, the remaining article will primarily focus on properties of the structure $\Pi$ and its computation. From a transformation of the form (1) which leaves the functional integral measure $[D\phi]$ invariant, the transformation of the average field $\phi_a(x)$ is

$$\phi_a'(x) = \phi_a(x) + \epsilon \langle F(x,\phi) \rangle \qquad (5)$$

$$[D\phi] \equiv [D\phi']$$

From (2),

$$\langle S[\phi'] \rangle = \langle S[\phi] \rangle + \epsilon \int d^d x \langle \frac{\delta S}{\delta \phi} \cdot F(x,\phi) \rangle \qquad (6)$$

As the average field $\phi_a(x)$ also belongs to the set of all field configurations $\Phi$, all measure preserving symmetries (invariant $[D\phi]$) of $S$ should correspond to a subset of the symmetries of $\Pi$ (vice-versa is not generally true). First, let us consider the case of continuous symmetries of S. Mathematically, $\Pi[\phi_a]$ is a functional of the average field $\phi_a$ such that for all $F(x,\phi)$,

$$\int d^d x \langle \frac{\delta S}{\delta \phi} \cdot F(x,\phi) \rangle = \int d^d x \frac{\delta \Pi}{\delta \phi_a} \cdot \langle F(x,\phi) \rangle \qquad (7)$$

In other words, for every measure preserving continuous symmetry of $S$, there is a corresponding continuous symmetry of $\Pi$. Let us derive the form of the probability distribution $P[\phi]$ for which (7) becomes an identity. Under all measure preserving field transformations of form (1), the functional integral $\int [D\phi]\, P[\phi]$ should remain invariant (as it is sum of probabilities of all fields). So, to ensure (7) is an identity, one can have

$$\int d^d x \frac{\delta P}{\delta \phi} = \int d^d x \frac{\delta \Pi}{\delta \phi_a} P[\phi] - \int d^d x \frac{\delta S}{\delta \phi} P[\phi] \qquad (8)$$

Integrating (8) with respect to $\phi$,

$$P[\phi] = \frac{1}{Z} e^{-S[\phi] + \int d^d x \frac{\delta \Pi}{\delta \phi_a}(x) \cdot \phi(x)} \qquad (9)$$

$$Z = \int [D\phi]\, e^{-S[\phi] + \int d^d x \frac{\delta \Pi}{\delta \phi_a}(x) \cdot \phi(x)} \qquad (10)$$

where $Z$ is the partition function which ensures $P[\phi]$ is a probability measure. It is then straightforward to realize that $\Pi[\phi_a]$ is the Legendre transform of $\ln(Z)$ (up to a constant independent of $\phi_a$)

$$\Pi[\phi_a] = \int d^d x \frac{\delta \Pi}{\delta \phi_a}(x) \cdot \phi_a(x) - \ln(Z) \qquad (11)$$

In Quantum Field theory (QFT), the functional $\Pi$ is the effective action with $S$ as the bare (microscopic) action and (7) is, then, a manifestation of the Slavnov-Taylor identities[9]. Although $\Pi$ is constructed so as to have a continuous symmetry corresponding to every measure preserving continuous symmetry for $S$, it is not directly evident if the correspondence would hold even for discrete symmetries of the bare functional $S$. To this end, consider a discrete invertible transformation which leaves the functional integral measure $[D\phi]$ invariant

$$\phi'(x) = \phi(x) + \Delta(\phi, x) \qquad (12)$$

$$\phi(x) = \phi'(x) + G(\phi', x)$$

$$[D\phi] \equiv [D\phi']$$

Then,

$$S[\phi'] = S[\phi] \;\Rightarrow\; \left.\frac{\delta S}{\delta \phi}\right|_{\phi'} \cdot \left(1 + \left.\frac{\partial \Delta}{\partial \phi}\right|_{\phi}\right) = \left.\frac{\delta S}{\delta \phi}\right|_{\phi} \qquad (13)$$

Let

$$H(x,\phi) = e^{\int d^d x \frac{\delta \Pi}{\delta \phi_a}(x) \cdot \Delta(\phi,x)} - \left(1 + \left.\frac{\partial \Delta}{\partial \phi}\right|_{\phi + G(\phi,x)}\right) \qquad (14)$$

From (12), (13) and (14),

$$\int d^d x \langle \frac{\delta S}{\delta \phi} \cdot H(x,\phi) \rangle$$
$$= \frac{1}{Z} \int d^d x \left[ \int [D\phi]\, e^{-S[\phi] + \int d^d x \frac{\delta \Pi}{\delta \phi_a}(x) \cdot (\phi(x) + \Delta(\phi,x))} \left.\frac{\delta S}{\delta \phi}\right|_{\phi} - \right.$$
$$\left. \int [D\phi]\, e^{-S[\phi] + \int d^d x \frac{\delta \Pi}{\delta \phi_a}(x) \cdot \phi(x)} \left.\frac{\delta S}{\delta \phi}\right|_{\phi} \cdot \left(1 + \left.\frac{\partial \Delta}{\partial \phi}\right|_{\phi + G(\phi,x)}\right) \right]$$
$$= \frac{1}{Z} \int d^d x \left[ \int [D\phi']\, e^{-S[\phi'] + \int d^d x \frac{\delta \Pi}{\delta \phi_a}(x) \cdot \phi'(x)} \left.\frac{\delta S}{\delta \phi}\right|_{\phi' + G(\phi',x)} - \right.$$
$$\left. \int [D\phi]\, e^{-S[\phi] + \int d^d x \frac{\delta \Pi}{\delta \phi_a}(x) \cdot \phi(x)} \left.\frac{\delta S}{\delta \phi}\right|_{\phi + G(\phi,x)} \right] = 0 \qquad (15)$$

Then from (7) and (15),

$$\int d^d x \langle \frac{\delta S}{\delta \phi} \cdot H(x,\phi) \rangle = \int d^d x \frac{\delta \Pi}{\delta \phi_a} \cdot \langle H(x,\phi) \rangle = 0 \qquad (16)$$

This shows that every measure preserving discrete symmetry of $S$ also leads to a corresponding continuous symmetry of $\Pi$.

Rewriting (11),

$$\Pi[\phi_a] = \ln\left( \frac{e^{\int d^d x \frac{\delta \Pi}{\delta \phi_a}(x) \cdot \phi_a(x)}}{\int [D\phi]\, e^{-S[\phi] + \int d^d x \frac{\delta \Pi}{\delta \phi_a}(x) \cdot \phi(x)}} \right) \qquad (17)$$

If one associates a) the average field $\phi_a$ and effective action $\Pi[\phi_a]$ with the layer of visible spins $v$ of an RBM like structure and b) $\phi$ and bare action $S[\phi]$ with the layer of hidden spins $h$ thereof, after suitable discretization, from (9) and (17)

$$\ln[p(v)] \equiv \ln[P[\phi_a]] = \Pi[\phi_a] - S[\phi_a] \quad (18)$$

This model (Fig. 1) suffers from a major issue computationally-

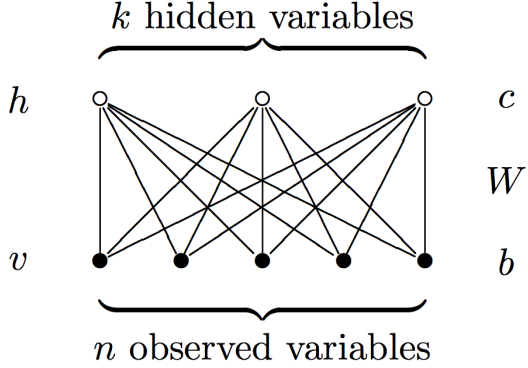

**Figure 1 | Graphical representation of the restricted Boltzmann machine.** Visible layer (observed variables) is comprised of units $v$ and the hidden layer (latent variables) is comprised of units $h$. The layers are fully connected, and the connections $W$ are bidirectional.

In general, the effective functional $\Pi$ is not solvable perturbatively in $\phi_a$ owing to divergent integrals (similar to effective action in QFT). In QFT, this brings in the need for exact non-perturbative renormalization group which aids one in studying the same physical system at different energy scales. Of particular interest, in this context, is the functional renormalization group which interpolates between the bare and effective action functionals[10]. Mathematically, the variation of the interpolating effective average action $\Pi_k$ with infrared energy (momentum) cutoff scale is described by the Wetterich flow equation (20). Effective average action $\Pi_k$ interpolates between effective action $\Pi$ ($k \to 0$) and bare action $S$ ($k \to \infty$).

$$e^{-\Pi_k[\phi_a]} = \int [D\chi] \exp\left(-S[\phi_a + \chi] + \int \frac{d^d p}{(2\pi)^d} \left[\widetilde{\frac{\delta \Pi_k}{\delta \phi_a(x)}}(-p) \cdot \tilde{\chi}(p) - \frac{1}{2}\tilde{\chi}^*(p) \cdot R_k(p) \cdot \tilde{\chi}(p)\right]\right) \quad (19)$$

Then in real space,

$$\frac{\partial \Pi_k[\phi_a]}{\partial k} = \frac{1}{2} \int d^d x \, d^d y \left[\frac{\partial \hat{R}_k(y-x)}{\partial k} \cdot \left(\frac{\delta^2 \Pi_k[\phi_a]}{\delta \phi_a(x) \delta \phi_a(y)}(x,y) + \hat{R}_k(x-y)\right)^{-1}\right] \quad (20)$$

Also[10],

$$\frac{\partial \Pi_k[\phi_a]}{\partial k} = \frac{1}{2} \int d^d x \, d^d y \left[\frac{\partial \hat{R}_k(y-x)}{\partial k} \cdot [\langle \phi(y)\phi(x)\rangle_k - \phi_a(y)\phi_a(x)]\right] \quad (21)$$

where $\tilde{f}(p)$ denotes the $d$-dimensional fourier transform of $f(x)$, $\hat{g}(x)$ denotes the inverse fourier transform of $g(p)$ and $R_k(p)$ is an IR regulator that provides a mass-like contribution to the bare action that suppresses the contributions of momenta $p \ll k$. Functional renormalization group flow equation (20) is exact and reproduces the effective action in the infrared limit ($k \to 0$) circumventing the divergences encountered in perturbative calculations[10]. Quite naturally, in order to mitigate the aforementioned issue with the RBM model, one might resort to incorporating the functional renormalization group idea into the proposed model. Rewriting the integral in (19) in terms of $\phi_\chi \equiv \phi_a + \chi$ and taking the logarithm,

$$\Pi_k[\phi_a] = \int \frac{d^d p}{(2\pi)^d} \left[\widetilde{\frac{\delta \Pi_k}{\delta \phi_a(x)}}(-p) \cdot \tilde{\phi}_a(p) + \frac{1}{2}\tilde{\phi}_a^*(p) \cdot R_k(p) \cdot \tilde{\phi}_a(p)\right] - \ln Z_k \quad (22)$$

$$Z_k = \int [D\phi_\chi] \exp\left(-S[\phi_\chi] + \int \frac{d^d p}{(2\pi)^d} \left[\widetilde{\frac{\delta \Pi_k}{\delta \phi_a(x)}}(-p) \cdot \tilde{\phi}_\chi(p) - \frac{1}{2}\left(\tilde{\phi}_\chi^*(p) \cdot \tilde{\phi}_\chi(p) - \tilde{\phi}_\chi^*(p) \cdot \tilde{\phi}_a(p) - \tilde{\phi}_\chi(p) \cdot \tilde{\phi}_a^*(p)\right) \cdot R_k(p)\right]\right) \quad (23)$$

One can identify (22) with (11) in the limit ($k \to 0$)[10]. As $S[\phi_a]$ is independent of $k$, (17) and (18) then allow us to associate every $\Pi_k[\phi_a]$ with the logarithm of the marginalized probability of a layer of spins in a cascading structure with each layer differing from the next, in the IR cutoff, by an infinitesimal amount $dk$. This lets us treat two consecutive layers in the cascade as an RBM and (20) shows how the effective average action for one layer depends on that of the next.

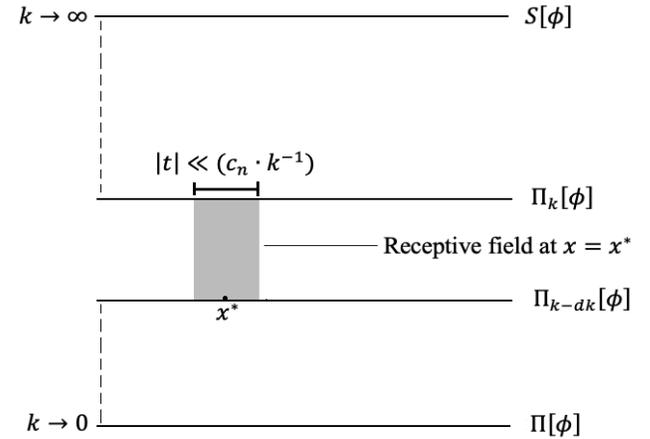

**Figure 2 | Graphical representation of the RG cascade.** The IR cutoff $k$ decreases as one moves down in the cascade recovering the full scale-independent effective functional (action) at the bottom.

Indeed, the action $S[\phi_a]$ serves as the logarithm of marginalized probability distribution of the uppermost layer ($k \to \infty$) and the exact effective action $\Pi[\phi_a]$ serves as the logarithm of marginalized probability distribution of the lowermost layer ($k \to 0$) in the resulting cascade (Fig. 2). As this cascading structure results from the exact non-perturbative renormalization group (20), computation of the effective functional $\Pi$ using this model is exact and free from integral divergences. This stacked model therefore circumvents the uncomputability, encountered in the single RBM model for computation, of the effective functional $\Pi$. One feature of the RG flow equation (20) viz., the IR regulator $R_k(p)$ has non-trivial implications for the cascade. As against the conventional RBM model, where all units in the hidden layer are connected to all units in the visible layer and vice-versa, the IR regulator $R_k(p)$ introduces a receptive field (a neighborhood) over the hidden spins (upper layer in the cascade) and the visible spins (lower layer in the cascade) for each spatial coordinate (Fig. 2). For the purposes of illustration, consider the following IR regulator $R_k(p)$ in a one-dimensional lattice

$$R_k(p) = \left(\frac{k}{p}\right)^{4n}, \quad n \in \mathbb{N} \quad (24)$$

where, $\mathbb{N}$ is the set of natural numbers. The regulator in (24) provides a smooth, yet sharp momentum cutoff at $p \sim k$ for large $n$. For $p \ll k$, the regulator $R_k(p)$ has a large value while for $p \gg k$, $R_k(p) \to 0$. Inverse fourier transform of (24) yields

$$\hat{R}_k(t) = \frac{k \cdot \pi}{(4n-1)!}(k \cdot t)^{4n-1} \cdot \text{sign}(t) \quad (25)$$

where, $\text{sign}(t)$ is the signum function. From (25), it can be seen that $R_k(t)$ also has a sharp spatial cutoff at $|t| \sim c_n \cdot k^{-1}$ for large n, where $c_n$ is a constant independent of $k$. For $|t| \ll (c_n \cdot k^{-1})$, $R_k(t) \to 0$ while for $|t| \gg (c_n \cdot k^{-1})$, $R_k(t)$ has a large value. From (25) and (20), it is easy to see that for every spatial coordinate $x$, only those spatial coordinates $y$ for which $|y - x| \lesssim (c_n \cdot k^{-1})$ would have a significant contribution to integral (20). It can be seen from (21), which is an alternative form of (20), how this generates a receptive field on the spins of consecutive layers (Fig. 2). This analysis is easily generalizable to higher dimensions as one can always select a regulator $R_k(p)$ which is separable in all dimensions and has form (24) in each dimension. The proposed cascade to compute the effective functional $\Pi$ from the bare functional $S$ does present a manifestation in conventional machine learning theory under suitable discretization viz., the convolutional deep belief network (CDBN). A CDBN comprises of stacked convolutional RBMs with an intermediate non-linear max-pooling layer for dimensionality reduction[11]. The hidden layer of one RBM in the stack serves as the visible layer of another (after max-pooling). Also, a convolutional RBM differs from a regular RBM in that the former introduces a receptive field over the hidden layer, for each spin in the visible layer, on which a convolution kernel (of the size described by the receptive field) acts to evaluate the probability of the visible spin. In the proposed cascade model, we have already established how one can treat every two consecutive layers as an RBM and we have also seen how the regulator $R_k(p)$ generates a receptive field over the consecutive layers. In light of the above, it can be seen how CDBN presents itself as a solution to the stochastic unsupervised learning problem by computing the effective functional $\Pi$ from the bare functional $S$ via the functional renormalization group flow (20). As against a conventional CDBN, however, no max-pooling layers are employed between two consecutive RBMs in the proposed cascade. The receptive field increases as one moves downwards in this CDBN ($k$ decreases) with a fully connected RBM at the bottom of the stack. CDBN aims to maximize $\ln(p(v))$ of configurations $v \in V$ of units in the lowermost layer, where $V$ is a subset of all possible configurations[11]. In the proposed cascade, this is analogously achieved by extremizing the effective functional $\Pi$ (supremum per definition of Legendre transform) as Schwinger functional $\ln(Z)$ is convex[12]. However, the bare functional $S$ is not a given in the case of a CDBN. Hence, the training procedure aims to compute $S$, the logarithm of the marginal probability of spins in the uppermost layer that maximizes the effective functional $\Pi$, the logarithm of the marginal probability of spins in the lowermost (visible) layer. One also needs to keep in mind that the RG step is exact only for an infinitesimal change in the cutoff scale $k$ and thus the proposed cascade is not directly realizable. At best, only a finite approximation of the proposed scheme is possible to implement.

The notion of symmetry is quite fundamental to both machine learning and physics. Previous attempts made to draw analogies based on this notion are either purely empirical or are plagued by assumptions which are not based on scientifically established principles[7]. The first half of the paper attempts to provide a solution to the stochastic unsupervised learning problem as formulating the effective functional $\Pi$ which encompasses all symmetries of stochastic data. This formulation is seen to be that of the effective action in QFT. The second half of the paper is devoted to computing $\Pi$ exactly, which suffers from integral divergences when a perturbative expansion is attempted (reminiscent of the direct uncomputability of effective action in QFT). A non-perturbative RG method, the functional renormalization group produces the exact effective action from the bare action in QFT and is therefore employed for computation of the effective functional Π. Form (17) and RG flow equation (20) then describe how functional RG can be performed by employing a cascade of RBMs with each layer differing from the next in the IR cutoff by an infinitesimal amount $dk$. Equation (25) illustrates how a suitable IR regulator $R_k(p)$ describes a receptive field over two consecutive layers in computation of (20). This receptive field increases as one moves downwards in the cascade ($k$ decreases) with a fully connected RBM at the bottom of the stack. As (20) is an exact equation, each stacked RBM computes the exact RG step. Although the conventional CDBN model provides a manifestation of the RG flow (20) for discrete lattice systems with layers of spins, the functional renormalization group is applicable to a wide class of physical systems (not necessarily discrete), whether in or out of equilibrium as against Ising models which only apply to lattice systems in thermal equilibrium. Functional RG has been used successfully in numerous problems in condensed matter and high energy physics (e.g., asymptotic safety). We present a manifestation of the proposed cascade model (under discretization) in the conventional machine learning theory viz., the CDBN and that the training procedure for CDBN computes the bare functional $S$, the logarithm of the marginal probability of spins in the uppermost layer that maximizes the effective functional Π, the logarithm of the marginal probability of the lowermost (visible) layer. In turn, this describes how a structure as complex as a convolutional deep belief network emerges as a solution to a stochastic unsupervised learning problem without making any assumptions about the underlying framework.